Title : Implementation of a Low-Cost, Distributed, Absolute Time Reference in an application dedicated to damage detection

Authors:   Louis-Marie Cottineau
           Vincent Le Cam
           Daniel-Marc Ducros

**ABSTRACT**

Typical structural health monitoring configuration implies sensors and supervisor installations connected by electric cable for communication. As done in other wireless projects, this one aim at reducing installation and maintenance costs by designing a wireless sensor network. One of the problem when designing wireless sensors, is data tagging: an event has to be correctly dated by many sensors. This problem increases when a precise time stamping is required.

What distinguishes this project is the implementation of algorithms that allows precise time stamping by the use of a standard protocol and the separation between measurements operations and communication task in order to make modular sensors.

**BACKGROUND AND ORIGIN : ACOUSTIC EMISSION**

**Global objective - Origin**

LCPC (Public Works Laboratory) develops new methods and instruments for monitoring civil engineering work, especially cable stayed or suspended bridges. More generally, the monitoring of breaks in cables is a great challenge for LCPC. These breaks relate to pre-stressed internal and external cables as well as those in bracing cables or suspension bridges.

In order to provide this surveillance a first generation detection system has already been developed and is a type of system called CASC: Control by Acoustic Surveillance of Cables. As its name indicates, cable monitoring is carried out by acoustic inspection. Cables of suspension bridges are generally multilayer single strand cables (4 to 5 mm wires) or assemblies of multilayer single strand cables positioned in parallel in the form of a layer or bundle.

This configuration means that when a wire breaks, its two ends re-anchor themselves both side from point of rupture.

___________

Louis-Marie Cottineau, Vincent Le Cam, Instrumentation Division, Laboratoire Central des Ponts et Chaussées, route de Bouaye, BP 4129, F 44341 Bouguenais Cedex, France.
Daniel-Marc Ducros, Engineering Sciences Division, L.C.P.C., route de Bouaye, BP 4129, F 44341 Bouguenais Cedex, France.

At this moment, an acoustic wave is generated in both direction from the point of rupture. Detection of this wave by various sensors makes it possible to locate the point of rupture and its magnitude. The detection of one (or more) point(s) of rupture together with location is information of strategic importance for managing the structure. Knowing the number of breaks and their location contributes to understanding the well-being of the structure.

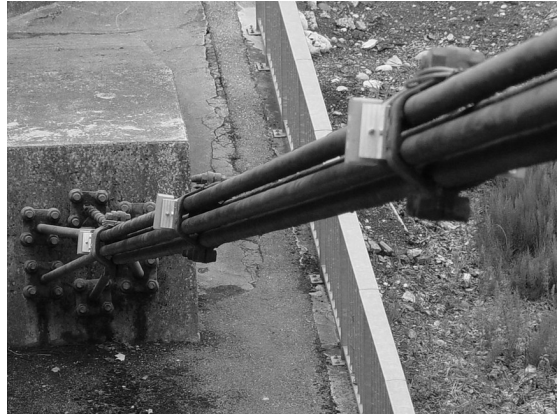

Figure 1 : initial version installed on Tancarville's bridge.

**Principle of operation**

Each sensor is composed of an accelerometer which uses the piezo-electric effect to produce electrical charges from the « fracture wave ». A variable detection threshold is set from a central PC, the supervisor. Sensors and the supervisor are connected together using a RS485 serial link. Typically, thresholds are set around 0.8g.

When the threshold of the nearest sensor is exceeded (the one that receives the wave first), it resets a time counter and sends a signal to the central PC and to the other sensors. Each sensor $i$ receiving this signal starts a time counter and records the maximum amplitude of the signal and the instant of detection $Ti$. All measurements are collected by the supervisor. The maximum amplitude over a pre-defined time (3 ms in the present system) after the threshold has been exceeded provides information on the energy released during rupture. Then, a software running in the supervisor PC determines the place where rupture occurred as showed in the figure below.

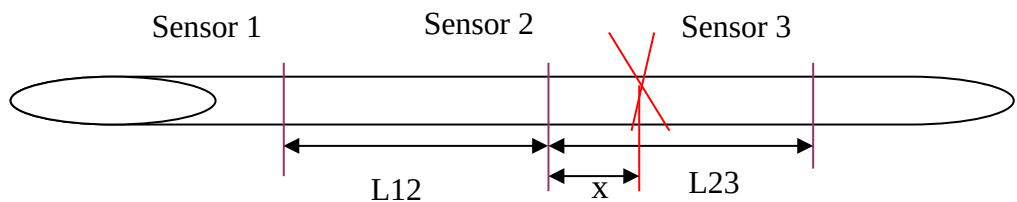

Figure 2 : principle of rupture detection.

First, the supervisor identifies sensors 1, 2 and 3. Sensors 2 and 3 are the two sensors that received the wave first. This gives the direction. Sensor 1 and sensor 2

(1)

allow to calculate the speed *v* of the wave. From these data it determines *X* by the calculation below :

$$v = L12 / \Delta t12$$
$$X = \frac{1}{2} * (L23 - v * \Delta t23) \qquad (2)$$

$\Delta tij$ = the difference in wave detection between sensor *i* and the sensor *j*.

**Results**

CASC system was able to detect and locate the actual ruptures of test cables, as was clear after these cables were removed from host and analyzed by other measures. The first generation of this system has been designed and put into production form and is now fitted to several French civil engineering constructions : Aquitaine Bridge in town of Bordeaux, bridge of Tancarville, bridge of Normandie for example.

A report is that the accuracy of measurement of *X* depends above all on the values of $\Delta t$. The accuracies established are of the order of ±15cm around *X*. A problem is the sensitivity of the CASC system to the noise (presence of waves that do not correspond to a rupture).

ASSUMPTIONS :
- Speed of the wave is constant (typical speed of the wave in the cable is around : 5000 m/s),
- Sensors are fixed with 10 m length between each two sensors,
- A cable is a dispersive entity.

**MOTIVATIONS FOR WIRELESS EXTENSIONS**

It has been decided to make the system more flexible and evolutionary by developing "wireless" sensors and with an option of a wire link. Sensors record the complete waveform of the signal (and no longer record only amplitudes) and make it possible to perform signal processing, frequency filtering and FFT types of processing in order to provide the user with more and more information. For example, this will help user to identify real rupture from noise (first system, based on amplitude detection was disturbed by noise).

The mean of sensor/supervisor communication must be faster, conforms to a protocol and has hight capability so that it can readily pass new data with a higher data rate. This applies to both wired and wireless modes. Sensors are designed to isolate effectively the processing unit from the communication unit.

In addition, and more generally, the project is also aiming at developing a solution which :
- Is "low cost": in terms of producing the sensor, supervisor and the means of communication,
- Has "development potential": sensors are readily adaptable to other types of instrumentation.

**SENSOR NETWORK ARCHITECTURE**

The need of wireless sensors is obvious for cost, facility, mobility reasons. In the field of wireless sensors, several solutions and products are developed since many years. What characterizes current projects is the use of many technologies emergence such as: micro-integration, low power consumption components, evolution of battery limitations, wireless protocols and, embarked API, etc.

For example, wireless sensors are developed for exploring and monitoring environment [1] or, in a more precise way, for detection and measurement of seismic waves in buildings [2], measurement of the air quality from dust sensors [3]. Other developments, like in the present project, integrate the will to offer a total measure and then, sensors have a modular measurement unit [4]; this allows an adaptation to acoustic domain, use of magnetometer, accelerometer etc.

Our approach was to study a solution so that the system is adaptable to another context or another data exchange protocol (between sensors and supervisor, between sensors,…) without "heavy" developments. We conceive a type of sensor which separates the treatment part of the communication part.

More than a wireless sensor, our wish is to develop a wireless network of sensors. We integrate the wireless TCP/IP technology offered by standards as 802.11b [5]. At the origin, those networks were mainly dedicated to office applications to design wireless computers, wireless printers, etc. It appeared interesting to us to use this technology in the field of civil engineering structures.

In order to arrive at a solution, the system is broken down in the following way:

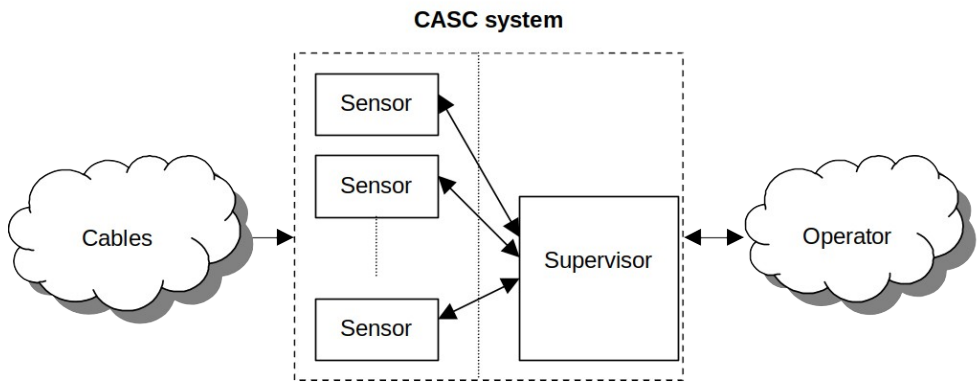

Figure 3 : system organization.

TABLE I : GLOBAL SOLUTION COMPOSITION

| Supervisor | |
|---|---|
| PC | Typical Personal Computer. Windows 98 runtime environment. |
| Access Point | Intel PRO/Wireless 5000 access point for the wireless network |
| Supervision software | Developed in C++ language supporting high level and intuitive interfaces. |
| **Communication based on the TCP / IP protocol** | |
| Ethernet (802.3) | For wired link, |

| 802.11b standard | For wireless link. |
|---|---|
| **Sensors, each consisting of** | |
| A DSP stage | Analog Device ADSP2991 for signal acquisition and processing. |
| WIFI module | Ubicom IP2022 module for wire and wireless communications |

The sensor consists of a small processor whose logical heart is the DSP that we program in order to receive and process the recorded signals (sampling at 250 kHz). The communication section of the sensor is provided by the TCP/IP stack contained in IP2022 module. The two modules are independent and maintain dialogue through common buffers. The «Supervisor –Sensors » system is a «Wireless Local Area Network » :

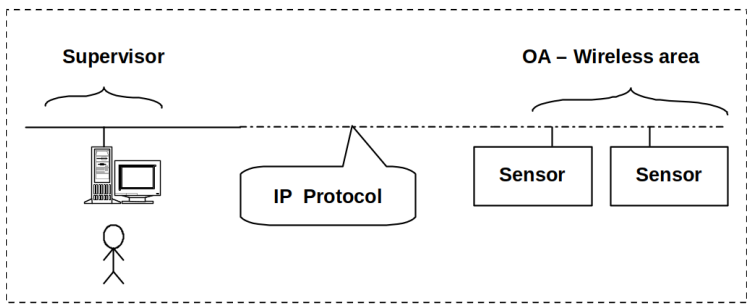

Figure 4 : wireless sensor network.

## DISTRIBUTION OF TIME SYNCHRONISATION AT LOW-COST

**Problems: The time synchronization of the sensors**

An event detected by the sensors is a wave exceeding a threshold value. The algorithms then aim at to digitize the shape of the signal from the instant of detection and to time-date samples. The accuracy with which the point of rupture is located depends on the accuracy with which the moment of detection is determined, and so sensors must time the same events using the same timebase.

The waves are moving at a velocity of around 5000 m/s and the required accuracy is ±15cm. In order to meet this degree of accuracy, the timing of the same event by 2 sensors $i$ and $j$ of the system must not differ more than 6µs.

The sensors are composed of electronic modules using quartz crystals whose drift is significant. Typically, a quartz crystal drifting at 50 ppm (parts per million) provides a clock that drifts 5000µs in 1s. In practice, the drift is lower but it still remains sufficiently large to justify the distribution of a reference time. There is therefore a need to distribute an absolute reference time to the sensors in order to synchronize them.

**Solutions studied**

**We have considered 3 solutions** :

1 – Each sensor is fitted with a module enabling it to "recover" absolute time; derived for example from a GPS (Global Positioning System) module. The problem here is that although accuracy is guaranteed, the cost of each sensor is high and cost grows linearly with the number of sensors. Solution 1 has therefore not been adopted.

2 – The supervisor recovers absolute time (with a GPS module, for example) and distributes it to the network of sensors. In this way, each moment the supervisor receives the time, each sensor can "re-time" recorded events by reference to the last reference received.

3 – Utilization of a low cost protocol on the LAN, dedicated to time synchronization and compatible with the TCP/IP network.

**Solution 2 : Ratiometric recalculation of events**

Because of the drift of quartz crystals due to temperature, aging, etc., solution 2 works but only for time resolution of the order of $10^{-5}$ s. Two measures of a same event collected by sensors $i, j$ at two moments $Tev_i, Tev_j$ can't be precisely compared because of sensor drift.

So, solution 2 had been reinforced by adding a ratiometric pinciple. A ratiometric calculation means that data that output from a system are evaluated, compared to a reference value. The concept developed has the following principle as shown in the figure below:

The supervisor, as «the leader» of the sensors network, transmits a synchronization frame named *Sync* at a period *T* to all the sensors.

Each sensor $i$ has an internal time counter $Ti$. When it receives *Sync*, each sensor saves its time counter $(Ti, Tj,…)$ and resets it, and then returns to the supervisor :

- The counter value saved $(Ti, Tj,…)$,
- The stored and timed event samples $(Tev_i, Tev_j,…)$.

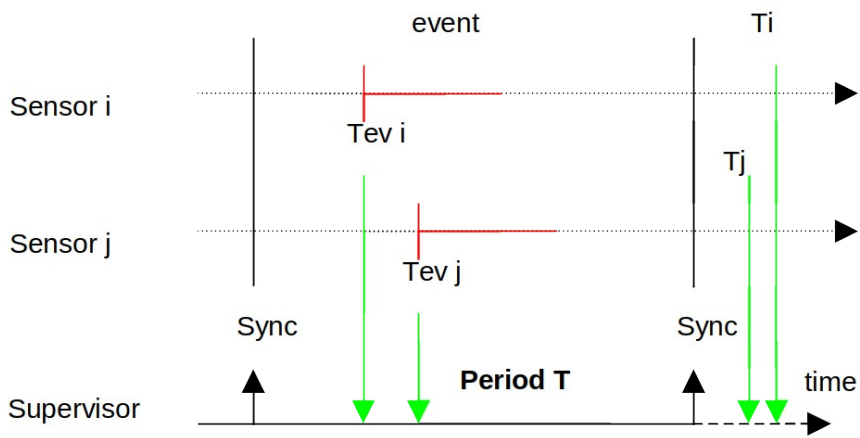

Figure 5 : synoptic of supervisor/sensors exchange.

The time difference between each sensor $i,j$ that receives the periodic synchronisation frame *Sync* is in agreement with the accuracy required as explained below in the "Time resolution" paragraph.

As this, the synchronisation period *T* can be used as a reference for all sensors. In this way, the supervisor can retime the events received during the period *T* in a ratiometric manner. Events can then be compared one against the others.

$$\text{Recalculated time: } Tevi' = Tevi * T / Ti \tag{3}$$

TIME RESOLUTION

In order to meet the required accuracy the synchronisation frame must be transmitted and taken into account for each sensor with a maximum discrepancy of 6 µs between each sensor. The message is sent on the WIFI network using the broadcast mode offered by the UDP/IP protocol.
The synchronization time of each sensor is the sum 2 times:
- The rf transmission velocity for a message, that is: $180*10^6$ m/s. The transmission time presents no problem since it is approximately the same for each sensor (in the absence of router and commutation, a LAN becomes deterministic). For a value of 6 µs, this allows an discrepancy of 1080 m between 2 sensors independently of the supervisor. This distance corresponds well to the context of a civil engineering structure*.
- The time of take note of the message. This is not zero but is virtually the same for all sensors. The sensors are identical and in the same state from the point of view of remote-communication since processing is de-correlated from the communication by the utilization respectively of a DSP module and a Wireless module. Furthermore, there is no OS such as Windows or Linux on board. Programming taking account of the sensor is deterministic.

* : In the case of a deterministic LAN, as the transit times can be determined, a possible development is to make provision at sensor level to take transit times into account, and in this way be able to deploy the network in other applications than engineering constructions (seismic applications for example).

The principle described above makes it possible not only to obtain accuracies of the order of a few µs but even better than that provided the drift from one sensor to another is small (a few tens of µs) with respect to the short resynchronisation duration (period *T*). This solution is an accurate and low-cost solution.

**Solution 3 : Utilization of a protocol**
The third solution that we test is the utilization of a «turn-key» protocol providing time synchronization on a TCP/IP network. We have not implemented this yet but we know that accuracies of the order of a microsecond or even a picosecond can be achieved if the components in the network and the network itself are designed in a deterministic manner. This is the case with our set-up where the sensors are designed with only one TCP/IP stack to dialogue and a LAN type network (no-commutation). Example: the Network Time Protocol V3 (6). Initial results from our tests enable us the following comparison:

TABLE II : INITIAL RESULTS

| Comparison domain | Solution 2: ratiometric calculation | Solution 3: turn-key protocol |
|---|---|---|
| Accuracy | Satisfactory, difficult to improve. | Satisfactory and capable of further development (Ipv6, NTP V4) … |
| Sensor consumption | Controlled by the synchronization period | Not controlled and higher. Depends largely on protocol utilized |
| Time supplied to the sensors | Relative to that of the supervisor without GPS (absolute if with GPS). | Absolutely that of the supervisor |
| Implementation | Needs development but the principle is valid for any system | Turnkey but valid in IP domain only. |

**Application**

The choice of a wireless protocol such as TCP/IP and the 802.11b standard allows a theoretical distance of 1 km between two sensors of the system. This technology implies the utilization of an access point to distribute the wireless network on radio waves. The distance between each sensor and a traditional access point connected to the supervisor is approximately from 100 m to 500 m. This configuration implies that all sensors and the supervisor are in an open field. To improve the area, repeaters and/or powerful antennas can be used.

The system in progress, without reinforcement, allows to equip bridges up to 300 m that correspond to a broad part of current bridges.

**CONCLUSION**

A wireless sensor system is developed. Now, the way that must be explored is the embedding of modules and the optimization of their power consumption.

The wireless solution developed offers more than cheaper installations and ease of maintenance, it also allows more mobile installations for statistical measurements for example. The use of the TCP/IP protocol provides better exchanges (for evolution and rate) between sensors and with the supervisor compared to conventional links (serial link or specific connection for examples). Even if the radio link if disturbed, typical 802.11b rate remains equals to 1 Mb/s.

The way sensors are designed allows them to be easily cabled if a wire connection is more appropriate (in a bridge for example). The benefit of the TCP/IP protocol if to be independent of the physical link. In the wire case, users have just to connect sensors by a traditional RJ45 cable.

The use of a specific DSP module for treatments separated from the wireless module provide a modular system. Sensors and supervisor can be programmed for other treatments or application and independently from the communication task. The use of the TCP/IP stack of the supervisor PC and the IP2022 module of the sensor make easy any system evolution. The above implemented solution allow sensor to have a low cost precise time synchronisation.

According to the number of industrialized sensors and selected technologies, the unit price should be included between a few tens and a few hundreds of dollars.

The wireless sensor network could equip smart structures and be used for other instrumentation field such as vibration measures using accelerometers, monitoring of bridges subjected with wind using pressure sensors, fatigue monitoring using deformation gaujes, thermics measures, chemical measures,…